\documentclass[12pt,twoside]{article}
\usepackage{fleqn,espcrc1}

\input{psfig}
\usepackage{graphicx}
\usepackage[figuresright]{rotating}
\usepackage{epsfig}

\newcommand{\be}{\begin{equation}}\newcommand{\ee}{\end{equation}}
\newcommand{\bea}{\begin{eqnarray}}\newcommand{\eea}{\end{eqnarray}}
\newcommand{\beaa}{\begin{eqnarray}}\newcommand{\eeaa}{\end{eqnarray}}
\newcommand{\ba}{\begin{array}}\newcommand{\ea}{\end{array}}
\newcommand{\bit}{\begin{itemize}}\newcommand{\eit}{\end{itemize}}
\newcommand{\ben}{\begin{enumerate}}\newcommand{\een}{\end{enumerate}}

\def\lab{\label}\def\lan{\langle}
\def\lf{\left}
\def\non{\nonumber}
\def\ran{\rangle}
\def\ri{\right}

\def\de{\delta}
\def\te{\theta}
\def\si{\sigma}\def\om{\omega}

\def\AB{{_{A,B}}}

\def\vec#1{{\bf #1}}

\begin{document}

\title{Phenomenology of flavor oscillations with non-perturbative effects 
from quantum field theory}

\author{Antonio Capolupo$^{a}$\thanks{E-mail: capolupo@sa.infn.it}
Chueng-Ryong Ji$^{b}$\thanks{E-mail: ji@ncsu.edu},
Yuriy Mishchenko$^{b}$\thanks{E-mail: ymishch@unity.ncsu.edu},
and Giuseppe Vitiello$^{a}$\thanks{E-mail: vitiello@sa.infn.it}\\
$^a$ Dipartimento di Fisica "E.R. Caianiello" and INFN,
Universit\`a di Salerno, I-84100
Salerno, Italy \\
$^b$ Department of Physics, North Carolina State University,
Raleigh, NC 27695-8202, USA}
\vspace{2mm}


\maketitle

\begin{abstract}
We analyze phenomenological aspects of the quantum field
theoretical formulation of meson mixing and obtain the exact
oscillation formula in the presence of the decay. This formula is
different from quantum mechanical formula by additional
high-frequency oscillation terms. In the infinite volume limit,
the space of the flavor quantum states is unitarily inequivalent
to the space of energy eigenstates.

\end{abstract}

\baselineskip=20pt


Quantum mixing of particles is among the most interesting and
important topics in Particle Physics\cite{chengli}. The Standard
Model involves quantum mixing in the form of Kobayashi-Maskawa
(CKM) mixing matrix {\cite{KM}}, a generalization of the original
Cabibbo mixing between $d$ and $s$ quarks {\cite{cabibbo}}. Also,
recently, convincing evidences of neutrino mixing have been
provided by Super-Kamiokande and SNO experiments
\cite{kamiokande,Kam,SNO,kamland,K2K}, thus suggesting neutrino
oscillations as the most likely resolution for the solar neutrino
puzzle \cite{solar} and the neutrino masses\cite{pascoli}. Since
the middle of the century, when the quantum mixing was first
observed in meson systems, this phenomenon has played a
significant role in the phenomenology of particle physics. Back in
1960s the mixing of $K^0$ and $\bar K^0$ provided an evidence of
CP-violation in weak interactions\cite{christenson} and more
recently the $B^0 {\bar B}^0$ mixing is used immensely to
experimentally determine the precise profile of CKM unitarity
triangle\cite{KM,cabibbo,jichoi}. Upgraded high-precision mixing
experiments in the meson sector would be vital to search for any
deviation from the unitarity of CKM matrix and thus put important
constraints on the new physics beyond the Standard Model. At the
same time, in the fermion sector, the discovery of neutrino mixing
and neutrino masses challenged our fundamental understanding of
CP-violation and, therefore, of the Standard Model itself.

Regarding the vanishing magnitudes of the expected new physics
effects (such as the unitarity violation in CKM matrix and/or
neutrino masses), it is imperative that the theoretical aspects of
the quantum mixing are precisely understood. In this direction, it
was noticed recently that the conventional treatment of flavor
mixing, where the flavor states are defined in the Fock space of
the energy-eigenstates, suffers from the problem of total
probability non-conservation \cite{BHV98}. This demonstrated that
the mixed states should be treated rather independently from the
energy-eigenstates. In fact, it was shown that the Fock space of
the mixed states is unitary inequivalent to the Fock space of the
energy-eigenstates and that the additional high-frequency term
must be present in the flavor oscillation formulas. Simpler
quantum mechanical result is reproduced only in the relativistic
limit of quantum field theory. In this respect, one may question
the magnitude of the field-theoretical effects and their
significance to the new physics in the mixing phenomena.

A significant research effort had been undertaken in the quantum
field theory of mixing
\cite{BHV98,BV95,Blasone:1999jb,binger,BCRV01,JM01,ABIV96,yBCV02,JM011}.
Still, the general theoretical results obtained therein cannot be
immediately applied to the phenomenologically interesting cases.
The mixing of particles and antiparticles in the meson sector
(e.g. $K^0-\bar K^0$, $B^0-\bar B^0$) requires specific
adjustments to the results obtained previously. Moreover, except
neutrinos, all known mixed systems are subject to decay and thus
the effect of particle life-time should also be taken into
account.

In this short note, we analyze the phenomenological aspects of the
nonperturbative field-theoretical effect in flavor mixing.
Specifically, we analyze the adjustments needed
for the general formulation in order to make applications for the known
systems. We also study the effect of the finite particle
life-time on the field-theoretical oscillation formula.
Finally we estimate the magnitudes of the nonperturbative corrections in
various systems and discuss the systems in which the field-theoretical
effect may be most significant.

In order to illustrate the field-theoretical method, we consider the
derivation of oscillation formulas for the case
of mixing of neutral bosons. We begin with the mixing relations
\bea\non
&&\phi_{A}(x) = \phi_{1}(x) \; \cos\te + \phi_{2}(x) \; \sin\te
\\[2mm] \lab{2.53}
&&\phi_{B}(x) =- \phi_{1}(x) \; \sin\te + \phi_{2}(x)\; \cos\te ,
\eea
where, generically, $\phi_A$ and $\phi_B$ are the fields associated
with the particles with given flavor and $\phi_{i}(x)$ are the "free"
fields with definite mass $m_{1,2}$. For the neutral particles, all
fields in Eq.(\ref{2.53}) are self-conjugate.
The Fourier expansions of the free fields $\phi_{1,2}$ and their
conjugate momenta $\pi_{1,2}$  are
\bea\lab{2.51}
\phi_{i}(x) = \int \frac{d^3k}{(2\pi)^{\frac{3}{2}}}
\frac{1}{\sqrt{2\om_{k,i}}} \lf( a_{{\bf k},i}\,
e^{-i \om_{k,i} t} + a^{\dag }_{{-\bf k},i}\, e^{i \om_{k,i} t}  \ri)
e^{i {\bf k}\cdot {\bf x}}
\eea
\bea\lab{2.52}
\pi_{i}(x) = i\,\int \frac{d^3 k}{(2\pi)^{\frac{3}{2}}}
\sqrt{\frac{\om_{k,i}}{2}} \lf( a^{\dag}_{{\bf k},i}\,
e^{i \om_{k,i} t} - a_{{-\bf k},i}\, e^{-i \om_{k,i} t} \ri)
e^{i {\bf k}\cdot {\bf x}}\,,
\eea
where $\om_{k,i}=\sqrt{{\bf k}^2+ m_i^2}$ and
the nonvanishing commutators are
$\lf[a_{{\bf k},i},a_{{\bf p},j}^{\dag} \ri]=
\de^{3}(\vec{k}-\vec{p}) \de_{ij}\, $
with $i,j=1,2$.

Following Ref.\cite{BV95}, we recast
Eq.(\ref{2.53}) into the form
\bea
\phi_{A}(x) = G^{-1}_\te(t)\; \phi_{1}(x)\; G_\te(t) \\[2mm]
\lab{2.53b}
\phi_{B}(x) = G^{-1}_\te(t)\; \phi_{2}(x)\; G_\te(t)
\eea
and similarly for $\pi_{A}(x)$ and $\pi_B(x)$.
Here, $G_\te(t)$  is the operator that furnishes
the representation of the mixing transformation
(\ref{2.53})
in the linear space of quantum fields and can be found as
\bea\lab{2.54neu}
G_\te(t) =exp\lf[-i\;\te \int d^{3}x
\lf(\pi_{1}(x)\phi_{2}(x) - \phi_{1}(x)\pi_{2}(x)\ri)\ri]\, .
\eea

In the finite volume, this is a unitary operator satisfying
$G^{-1}_\te(t)=G_{-\te}(t)=G^{\dag}_\te(t)$ which may be written as
\bea\lab{2.55neu} G_\te(t) = \exp[\te S(t)] ~, \eea
with
\bea\lab{2.58neu} S(t)=\int d^3 k \lf( U^*_{{\bf k}}(t) \,
a_{{\bf k},1}^{\dag}a_{{\bf k},2} - V_{{\bf k}}^{*}(t) \,
a_{{\bf k},1}a_{{\bf k},2} + V_{{\bf k}}(t) \,
a_{{\bf k},1}^{\dag}a_{{\bf k},2}^{\dag} - U_{{\bf k}}(t) \,
a_{{\bf k},1}a_{{\bf k},2}^{\dag}
\ri).
\eea
The coefficients
$U_{{\bf k}}(t)\equiv |U_{{\bf k}}| \;
e^{i(\om_{k,2}-\om_{k,1})t}$ and $V_{{\bf k}}(t)
\equiv |V_{{\bf k}}| \;e^{i(\om_{k,1}+ \om_{k,2})t}$
are the coefficients of Bogoliubov transformation
defined by
\bea
&&|U_{{\bf k}}|\equiv \frac{1}{2}
\lf( \sqrt{\frac{\om_{k,1}}{\om_{k,2}}}
+ \sqrt{\frac{\om_{k,2}}{\om_{k,1}}}\ri) ~,
\;\;\;\;\;\;
|V_{{\bf k}}|\equiv  \frac{1}{2} \lf(
\sqrt{\frac{\om_{k,1}}{\om_{k,2}}}
- \sqrt{\frac{\om_{k,2}}{\om_{k,1}}} \ri).
\eea
They satisfy the unitarity relation
\bea\lab{2.60} &&|U_{{\bf k}}|^{2}-|V_{{\bf k}}|^{2}=1\,, \eea
and thus can be put in the form
$|U_{{\bf k}}|\equiv \cosh\xi^{\bf k}_{1,2}$ and
$|V_{{\bf k}}|\equiv \sinh \xi^{\bf k}_{1,2}$ with
$\xi^{\bf k}_{1,2}= \frac{1}{2} \ln\frac{\om_{k,1}}{\om_{k,2}}$.

The mixing transformation also induces a $SU(2)$ coherent state
structure on the quantum states \cite{Per} and the vacuum state
given by \bea\label{2.61} |0(\te, t) \ran_\AB \equiv
G^{-1}_\te(t)\; |0 \ran_{1,2}\,. \eea We refer to state $|0(\te,
t) \ran_\AB$ as the "flavor" vacuum for the mixed fields
$\phi_{A,B}$ \cite{BV95}.

Let us now consider the Hilbert space of the flavor fields
at a given time $t$, say $t=0$.
It is useful to define $|0 (t)\ran_\AB\equiv|0(\te,t)\ran_\AB$
and $|0 \ran_\AB\equiv|0(\te, t=0)\ran_\AB$.
In the infinite volume limit the flavor and the mass vacua are
orthogonal
\cite{BCRV01}.
We observe that the orthogonality disappears when $\te=0$ and/or
$m_1=m_2$, which is consistent with the fact that in both cases there
is
no mixing.
For the flavor fields $\phi_{A,B}$ we then introduce the
annihilation/creation
operators $a_{{\bf k},A}(\te ,t) \equiv G^{-1}_\te(t) \;
a_{{\bf k},1}\;G_\te(t)$ such that $a_{{\bf k},A}(\te ,t) |0(t)
\ran_\AB =0$. For simplicity, we will use notation
$ a_{{\bf k},A}(t) \equiv a_{{\bf k},A}(\te ,t)$.
Explicitly, we have
\bea \label{2.62aneu}
a_{{\bf k},A}(t)&=&\cos\te\;a_{{\bf k},1}\;+\;\sin\te\;
\lf( U^*_{{\bf k}}(t)\; a_{{\bf k},2}\;+\;
V_{{\bf k}}(t)\; a^{\dag}_{{\bf k},2}\ri)\, ,
\\
a_{{\bf k},B}(t)&=&\cos\te\;a_{{\bf k},2}\;-\;\sin\te\;
\lf(U_{{\bf k}}(t)\; a_{{\bf k},1}\;- \; V_{{\bf k}}(t)\;
a^{\dag}_{{\bf k},1}\ri)\,.
\eea

We are now in position to address the question of flavor
oscillations for neutral bosons.
We note that the oscillating observable should be specified properly here,
because for the neutral fields all conventional charges
are trivially zero($Q_{A,B} \equiv 0$). As shown in 
\cite{Blaspalm},
however, the momentum operator for the mixing of neutral fields
may be analogous to the charge operator for charged fields.
In fact, if we define momentum operator for free fields by
\bea {{\cal P}}_{i} =
\int d^3 x [\pi_{i}(x)\nabla\phi_{i}(x)]= \int d^3 k
\frac{{{\bf k}}}{2} \Big(a^{\dag}_{{\bf k},i} a_{{\bf k},i}-
a^{\dag}_{{-\bf k},i} a_{{-\bf k},i}\Big)
\eea
and, similarly, for mixed fields,
\bea {{\cal P}}_{\sigma} =\int d^3 x
[\pi_{\sigma}(x)\nabla\phi_{\sigma}(x)]= \int d^3 k \frac{{{\bf
k}}}{2} \Big(a^{\dag}_{{\bf k},\sigma}(t) a_{{\bf k},\sigma}(t)-
a^{\dag}_{{-\bf k},\sigma}(t) a_{{-\bf k},\sigma}(t)\Big),
\eea
then we can show that the total momentum is conserved in time:
${{\cal P}}_{A}(t)+{{\cal P}}_{B}(t)={{\cal P}}_{1}+{{\cal P}}_{2}
={{\cal P}}$.
The expectation value of the momentum operator at $t\neq 0$,
normalized to its initial value, is given by
\bea\label{simomenta}
 {{\cal P}}_{{\bf k},\si}(t)&\equiv& \frac{{}_\AB\lan {
a}_{{\bf k},A} | {\cal P}_\si(t) | {a}_{{\bf k},A}\ran_\AB
\,}{{}_\AB\lan { a}_{{\bf k},A} | {\cal P}_\si(0) | {a}_{{\bf
k},A}\ran_\AB \,} =\, \lf|\lf[{a}_{{\bf k},\si}(t),
{a}^{\dag}_{{\bf k},A}(0) \ri]\ri|^2 \; - \;
\lf|\lf[{a}^\dag_{-{\bf k},\si}(t), {a}^{\dag}_{{\bf k},A}(0)
\ri]\ri|^2 \quad, \\ \si=A,B\, . \eea
Explicitly, \bea
\label{Amomentum}
 {\cal P}_{{\bf k},A}(t)&=&
1 - \sin^{2}( 2 \theta) \lf[ |U_{{\bf k}}|^{2} \; \sin^{2}
\lf( \frac{\omega_{k,2} - \omega_{k,1}}{2} t \ri) -|V_{{\bf
k}}|^{2} \; \sin^{2} \lf( \frac{\omega_{k,2} + \omega_{k,1}}{2} t
\ri) \ri] \, ,
\\[4mm]
{\cal P}_{{\bf k},B}(t)&=& 
\sin^{2}( 2 \theta)\lf[ |U_{{\bf k}}|^{2} \; \sin^{2} \lf(
\frac{\omega_{k,2} - \omega_{k,1}}{2} t \ri) -|V_{{\bf k}}|^{2} \;
\sin^{2} \lf( \frac{\omega_{k,2} + \omega_{k,1}}{2} t \ri) \ri].
\label{Bmomentum}
\eea

\begin{figure}
\centering
\begin{minipage}[c]{0.4\hsize}
\epsfig{width=\hsize,file=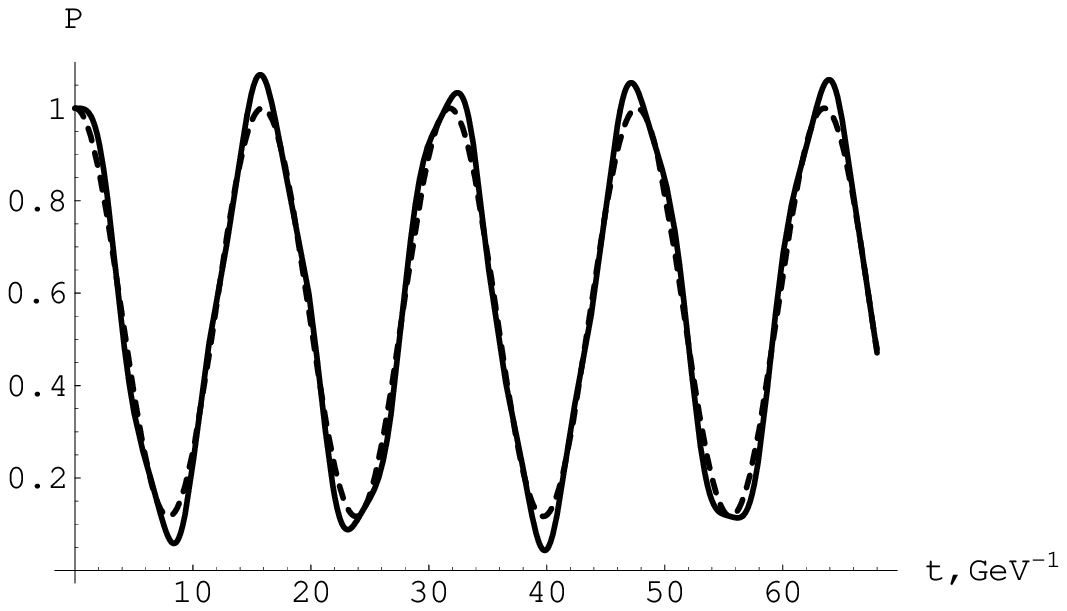}
\end{minipage}
\hspace*{0.5cm}
\begin{minipage}[c]{0.4\hsize}
\epsfig{width=\hsize,file=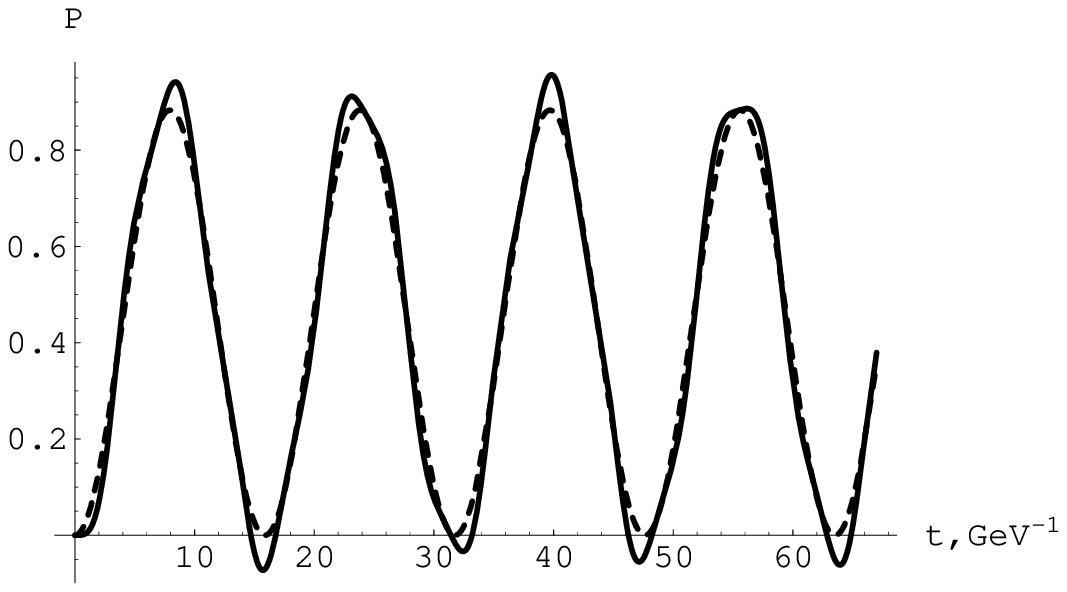}
\end{minipage}
\caption{Relative population densities ${\cal P}_A$ (left) and
${\cal P}_B$ (right) as the function of t for $k=0.1 GeV$ in
$\eta-\eta'$ system ($m_\eta = 549 MeV, m_\eta^\prime = 958 MeV$
and $\theta \approx -54^\circ$ \protect\cite{JM01}).
Solid line - QFT result, dashed line - QM result.}\label{fig1}
\end{figure}

Eqs.(\ref{Amomentum})-(\ref{Bmomentum}) are the flavor oscillation
formulas for the neutral mesons, such as $\eta-\eta'$,
$\phi-\omega$ etc. By the definition of the momentum operator,
Eqs.(\ref{Amomentum})-(\ref{Bmomentum}) are the relative population
densities of flavor particles in the beam. As an example, ${\cal
P}_A$ and ${\cal P}_B$ for the $\eta-\eta^\prime$ system are
plotted in Fig.\ref{fig1} as the function of time.

Still, for systems like $K_0-\bar K_0$ some
more care needs to be taken.
Specifically, in $K^{0}-\bar{K}^{0}$ mixing, $K^0$ may not be treated
as neutral since $K^0\neq \bar K^0$.
Of course, this is not the case of mixing of two different
charged particles either. Rather, the particle here is mixed with its
antiparticle.
To establish a connection with our previous discussion,
it is important to identify the mixed degrees of freedom
properly.
Note that in $K^{0}-\bar{K}^{0}$ mixing there are three distinct
modes,  namely the strange eigenstates $K^{0}-\bar{K}^{0}$, the
mass eigenstates $K_L-K_S$ and the CP eigenstates $K_1-K_2$. Each
pair can be written as a linear combination of the other ones,
e.g.
\begin{equation}\label{yuri01}
\begin{array}{c}
K_1=\frac{1}{\sqrt{2}}(K^0 + \bar K^0),\quad
K_2=\frac{1}{\sqrt{2}}(K^0 - \bar K^0); \cr
K^0=\frac{e^{i\delta}}{\sqrt{2}}(K_L+K_S), \quad
\bar K^0=\frac{e^{-i\delta}}{\sqrt{2}}(K_L-K_S); \cr
K_1=\frac{1}{\sqrt{1+|\epsilon|^2}}(K_S+\epsilon K_L), \quad
K_2=\frac{1}{\sqrt{1+|\epsilon|^2}}(K_L+\epsilon K_S);
\end{array}
\end{equation}
with $e^{i\delta}$ being a complex phase and $\epsilon=i\delta$
being the imaginary CP-violation parameter. In a sense,
$K^{0}-\bar{K}^{0}$ are produced as strange eigenstates, propagate
as mass eigenstates $K_L, K_S$ and decay as CP-eigenstates
$K_1, K_2$.

The mass eigenstates
$K_L$ and $K_S$ are defined as the +1 and -1 CPT eigenstates,
respectively,
so that they can be represented in terms of self-adjoint scalar fields
$\phi_1, \phi_2$ as
\begin{equation}
K_L=\phi_1, \quad K_S = i\phi_2.
\end{equation}
Therefore the mixing in this system is similar to
the case of neutral fields with
{\em complex} mixing matrix. Since the complex mixing matrix in SU(2)
can be always transformed into the real one by suitable redefinition of
the field phases, which would not affect the expectation
values, the mixing in this case is still equivalent to the mixing of
neutral fields.
The oscillating observables may be that of the strange charge (in
the system $K^0$ and $\bar K^0$ taken as flavor A and B,
respectively) with the trivial mixing angle $\theta=\pi/4$ from
Eq.(\ref{yuri01}). Phenomenologically relevant, however, is the
oscillation of CP-eigenvalue which determines the ratio of
experimentally measured $\pi\pi$ to $\pi\pi\pi$ decay rates.
CP-oscillations are given in terms of $K_1$ and $ K_2$ flavors
with small mixing angle $\cos \theta =1/\sqrt{1+|\epsilon|^2}$.

The particle decay is taken in account by inserting by hand, as
usually done, the factor $e^{-\gamma t}$ in the annhilation
(creation) opeartors: $ a_{{\bf k},i} \rightarrow a_{{\bf
k},i}e^{-\frac{\gamma_{i}}{2}t}. $ Then, the oscillation formulas
can be written as \bea \non {\cal P}_{{\bf k},A}(t)&=&
\lf|\lf[a_{{\bf k},A}(t), a^{\dag}_{{\bf k},A}(0) \ri]\ri|^2 \; -
\; \lf|\lf[a^\dag_{-{\bf k},A}(t), a^{\dag}_{{\bf k},A}(0)
\ri]\ri|^2
\\[2mm] \label{Achargedecay}
&=& \lf(\cos^{2}\theta e^{-\frac{\gamma_{1}}{2}t}+\sin^{2}\theta
e^{-\frac{\gamma_{2}}{2}t} \ri)^{2}
\\[2mm] \non
&-& \sin^{2}(2\theta)e^{-\frac{\gamma_{1} + \gamma_{2}}{2}t} \lf[
|U_{{\bf k}}|^{2} \; \sin^{2} \lf( \frac{\omega_{k,2} -
\omega_{k,1}}{2} t \ri) -|V_{{\bf k}}|^{2} \; \sin^{2} \lf(
\frac{\omega_{k,2} + \omega_{k,1}}{2} t \ri) \ri] \, ,
\\[4mm]  \non
{\cal P}_{{\bf k},B}(t)&=& \lf|\lf[a_{{\bf k},B}(t),
a^{\dag}_{{\bf k},A}(0) \ri]\ri|^2 \; - \;
\lf|\lf[a^\dag_{-{\bf k},B}(t), a^{\dag}_{{\bf k},A}(0) \ri]\ri|^2
\\[2mm] \label{Bchargedecay}
&=& \sin^{2}( 2 \theta )
\lf(\lf[\frac{e^{-\frac{\gamma_{1}}{2}t}-e^{-\frac{\gamma_{2}}{2}t}}{2}
\ri]^{2} \right.
\\[2mm]  \non
&+&\left. e^{-\frac{\gamma_{1} + \gamma_{2}}{2}t}
\lf[ |U_{{\bf k}}|^{2} \;
\sin^{2} \lf( \frac{\omega_{k,2} - \omega_{k,1}}{2} t \ri) -
|V_{{\bf k}}|^{2} \; \sin^{2} \lf( \frac{\omega_{k,2} +
\omega_{k,1}}{2} t \ri) \ri] \right) . \eea
We note the difference between these oscillation formulas
and the quantum mechanical Gell-Mann--Pais formulas. Essentially,
the quantum field theoretic corrections appear as
the additional high-frequency oscillation terms.

\begin{figure}
\begin{center}
\epsfig{width=250pt, file=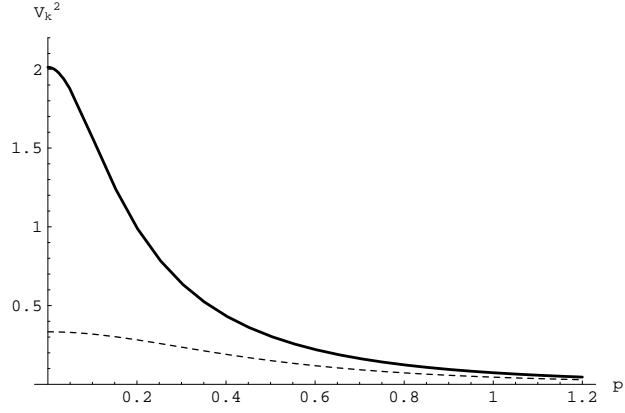} \caption{The bosonic
condensation density $|V(p,a)|^2$ as a function of $p$ for
$a=0.98$ (solid line) and $a=0.8$ (dashed line). }\label{fig4}
\end{center}
\end{figure}

In all of field-theoretical derivations (See
Eq.(\ref{Amomentum})-(\ref{Bchargedecay})), the field-theoretical
effect (or the high-frequency oscillation term) is proportional to
$|V_{\bf k}|^2$. In estimating the maximal magnitude of this term,
it is useful to write $|V_{{\bf k}}|^{2}$ in terms of the
dimensionless momentum $p\equiv\sqrt{\frac{2 |{\bf
k}|^2}{m_1^2+m_2^2}}$ and the dimensionless parameter $a\equiv
\frac{m_2^2-m_1^2}{m_1^2 +m_2^2}$ so that \bea |V(p,a)|^2 & =&
\frac{p^2 +1}{2\sqrt{(p^2 + 1)^2 - a^2}} -\frac{1}{2} . \eea
As shown in Fig.\ref{fig4},
$|V_{\bf k}|^2$ is maximal at $p=0$ ($|V_{max}|^2 = \frac{(m_1
-m_2)^2}{4 m_1 m_2}$) and goes to zero for large momenta (i.e. for
$|{\bf k}|^2\gg\frac{m_1^2 +m_2^2}{2}$).
The optimal observation scale for field-theoretical effect in
meson mixing, therefore, is $k=0$ and the maximal correction is of
the order of $|V|^2\sim \frac{\Delta m^2}{m^2}$. It is
straightforward to find that relative field-theoretical effect in
$K^0 - \bar K^0$, $D^0 - \bar D^0$, $B^0 - \bar B^0$ and $B^0_s -
\bar B^0_s$ is very small and generally does not exceed
$10^{-26}$. At the same time, for $\omega-\phi$ and $\eta-\eta'$
field-theoretical corrections may be as large as $5\% -20\%$,
respectively, and thus one needs to be careful about taking them
into account should these systems ever be used in some sort of
mixing experiments.

\begin{figure}
\begin{center}
\epsfig{width=250pt,file=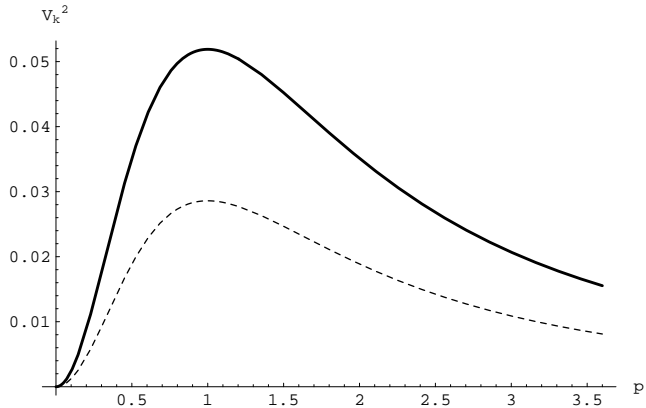}
\end{center}
\caption{The fermionic condensation density $|V(p,a)|^2$ as a
function of $p$ for $a=0.98$ (solid line) and $a=0.5$ (dashed
line). }\label{fig5}
\end{figure}

We can employ the similar method in the fermion sector.
Since neutrinos are stable, no additional adjustments are necessary to
the known results \cite{BV95}.
We can write
the field-theoretical correction amplitude $|V_{{\bf k}}|^{2}$
as a function of the dimensionless momentum
$p=\frac{|{\bf k}|}{\sqrt{m_1 m_2}}$ and dimensionless
parameter $a= \frac{m_2^2-m_1^2}{m_1 m_2}$, as follows,
\bea |V(p,a)|^2 & =& \frac{1}{2}(1-\frac{p^2+1} {\sqrt{(p^2 + 1)^2
+ a p^2}}) ~.\label{Vpa} \eea
From Fig.\ref{fig5} we see that the effect is maximal when $p=1$
($|V_{max}|^2 \approx \frac{(m_1 -m_2)^2}{16 m_1 m_2}$) and
$|V|^2$ goes to zero for large momenta (i.e. for $|{\bf k}|^2\gg
\frac{m_1^2 +m_2^2}{2}$) as $|V|^2
\approx \frac{\Delta m^2}{4 k^2}$.

Since we do not know yet the values of neutrino masses, we cannot
properly specify the optimal scale for observation of field-theoretical
effect in this sector. However, certainly this scale cannot be much
larger
than a fraction of eV.
So far the experimentally observed neutrinos are always extremely
relativistic
and, therefore, the value of $|V|^2$ may be estimated as
$|V|^2 \sim \frac{\Delta m^2}{k^2} \sim 10^{-18}$. Only
for extremely low energies (like those in
neutrino cosmological background) the field-theoretical corrections
might be
large and account for few percent.
In this connection, we observe that the non-perturbative field
theory effects, in spite of the small corrections they induce in
the oscillation amplitudes, nevertheless they may contribute in a
specific and crucial way in other physical contexts or phenomena.
An example of this is provided by the recent result
 \cite{costcosm} which shows
that the mixing of neutrinos may specifically contribute to the
value of the cosmological constant exactly because of the
non-perturbative effects expressed by the non-zero value of
$|V_{\bf k}|^2$.

To summarize, in this note we considered phenomenological
aspects of the quantum field theoretical formalism for
spin-zero boson-field mixing.
A crucial point in our analysis is the disclosure of the
fact that the space for the mixed field states is unitarily
inequivalent to the state space where the unmixed field operators are
defined. This is a common feature with the QFT structure of
mixing, which has recently been established.
The vacuum for the mixed fields turns out to be a generalized
$SU(2)$ coherent state.

We have estimated the magnitude of the field-theoretical effect in
known mixed systems. We found that for most of known mixed systems
both in meson and neutrino sectors this effect is negligible. Only
in strongly mixed systems, such as $\omega-\phi$ or $\eta-\eta'$,
or for very low-energy neutrino effects the corrections may be as
large as $5\% -20\%$ and thus additional attention may be needed
if these systems can be used in oscillation experiments.
The non-perturbative vacuum effect is the most prominent when the
particles are produced at low momentum.

This work was supported in part by a grant from the U.S. Department
of Energy(DE-FG02-96ER 40947) and the National Science Foundation
(INT-9906384). We also thank MURST, INFN and the ESF Program COSLAB for
partial financial support.

\vspace{.5cm}
%

\end{document}